\documentclass[twocolumn,showpacs,amsmath,amssymb]{revtex4}
\usepackage{amssymb}
\usepackage[dvips]{graphicx}
\begin{document}

\title{Quantum magnetic oscillations and angle-resolved photoemission from impurity bands in cuprate superconductors}
\author{A. S. Alexandrov$^*$}

\affiliation{Instituto de Fisica Gleb Wataghin/DFA,
Universidade Estadual de Campinas-UNICAMP 13083-859,
Brasil\\}

\begin{abstract}
Present-day  angle-resolved photoemission
  spectroscopy (ARPES) has offered a tremendous
advance in the understanding of  electron energy spectra in
cuprate superconductors and some related compounds. However, in high magnetic field, magnetic quantum oscillations  at
low temperatures  indicate the existence of small electron (hole)
Fermi  pockets  seemingly missing in ARPES of  hole (electron) doped cuprates. Here ARPES and quantum oscillations  are reconciled in the framework of an impurity band in the charge-transfer   Mott-Hubbard insulator.
\end{abstract}


\maketitle
ARPES of  cuprate superconductors \cite{dam}
proved to be particularly instrumental in modeling the electron energy spectrum of these charge transfer Mott-Hubbard insulators  showing that below optimal doping the single-particle Fermi
surface is reduced to four disconnected peculiar-shaped spots.
While on the overdoped side of the
phase diagram a large Fermi surface
is expected,   a few holes doped
into the insulator would naturally give rise to four hole "nodal Fermi
pockets" with a small area  proportional to the
doping on the underdoped side of the phase diagram. Another possibility is
a truncation of the large hole Fermi surface
giving rise to four "nodal Fermi arcs" due to a highly anisotropic quasiparticle life-time  and/or  a d-wave-like pseudogap.

Remarkably,  magnetic quantum
oscillations (MQO) in kinetic  and magnetic  response
functions of oxygen-ordered ortho-II YBa$_2$Cu$_3$O$_{6.5}$ (YBCO6.5) and of some other cuprates \cite{undo} revealed  small \textit{electron} Fermi pockets, rather than hole pockets or arcs
 seemingly in disagreement with ARPES results \cite{dam}.

 A number of Fermi surface reconstructions \cite{saw} and non-Fermi-liquid models, including our modulated vortex lattice scenario \cite{asavor}, have tempted to account for the nature of these unusually slow MQO. Further  careful experiments have found  the Zeeman splitting in MQO, which separates spin-up and spin-down contributions, indicating that  electrons in  cuprates behave as nearly free spins, which rules out most of the reconstruction and non-Fermi liquid scenarios \cite{proust}. This observation as well as excellent fits of MQO data  with field-independent oscillation frequencies  support the view \cite{undo} that MQO are a signature of the true zero-field normal state. This view is also supported by the observation of similar slow oscillations in  electron-doped cuprates Nd$_{2-x}$Ce$_x$CuO$_4$ \cite{kar}, where  due to their lower critical fields,  the normal state is reliably accessed for any doping level.
 In contrast with the \textit{electron} pockets in \textit{hole} doped cuprates,
MQO  reveal  small \textit{hole} pockets around certain doping level $x\approx 0.165$ in \textit{electron} doped cuprates \cite{kar2}.

Until recently these  two
 different measurement techniques, ARPES and MQO, were carried out on different
materials.
Aiming to resolve the outlined puzzle  Hossain et al. \cite{hos}  succeeded to
control the surface doping and follow the evolution of ARPES from the overdoped to the underdoped regime through an in situ deposition
of potassium atoms on cleaved YBCO.  Ref. \cite{hos} did not find any  sign of the electron pockets in the ARPES
data from underdoped YBCO, nor any sign of extra zone folding
due to the kind of density wave instabilities that might give
rise to such a Fermi surface reconstruction.
 Hossain et al. argued that if any pocket had to be postulated on the basis of their
ARPES data, the most obvious possibility would be that the Fermi
arcs are in fact   nodal Fermi pockets.  However, these are hole, not electron
pockets.

Apparently  without a detailed microscopic theory
both  ARPES and MQO data remain a mystery. Here I reconcile  these two rather precise techniques by introducing an impurity band in the doped charged transfer Mott-Hubbard insulator. Small electron (hole) pockets in the impurity band account for MQO observed in hole (electron) doped cuprates \cite{undo,kar}, and for the "half-moon"  spots  in ARPES \cite{hos}.

 Cuprate superconductors  are strongly correlated
electron systems where the \textit{ab-initio} local-density-approximation (LDA)  fails especially in the undoped regime. Fortunately,
 adding the on-site Coulomb repulsion U to the LDA analysis within  the  LDA+U algorithm \cite{ani} or using LDA plus the tight-binding (cluster) approximation (LDA+GTB algorithm \cite{ovc}), one can reproduce the correct magnetic
ground state and the charge-transfer gap in parent insulators such as La$_2$CuO$_4$ (LCO). The latter and some other schemes \cite{mar}  for the electronic structure found the charge-transfer gap also in the paramagnetic state   of cuprates, pointing to a persistent Mott physics also at finite doping \cite{ovc2}.

Different from the reconstruction models, proposed so far, I suggest that YBCO6.5 is the Mott-Hubbard insulator where the Fermi level is pinned within the   charge-transfer gap as in its  parent insulator YBCO6.0 \cite{saw}. This assumption is supported by  experiments
with ultrathin insulating La$_2$CuO$_4$ and doped superconducting La$_{2-x}$Sr$_x$CuO$_4$ (LSCO) layers \cite{boz} and by earlier optical spectroscopies of  YBCO \cite{taliani,mic}
 indicating that doped electronic states appear within the charge-transfer gap. These states, localized in band-tails by disorder,  readily account
for sharp   "quasiparticle" peaks and a rapid loss of their intensities in some directions of the Brillouin zone observed in ARPES   of lightly doped LSCO \cite{alearpes}. The same band-tail model explains  two energy scales, their temperature and doping dependence, the asymmetry and inhomogeneity of tunneling spectra of cuprate superconductors \cite{aletun}. Pinning the Fermi level   within the impurity band-tail is also compatible with
the insulating-like low-temperature normal state resistivity and  many other kinetic properties of lightly doped cuprates \cite{alebook,alearpes,neto}.

\begin{figure}
\begin{center}
\includegraphics[angle=-0,width=0.40\textwidth]{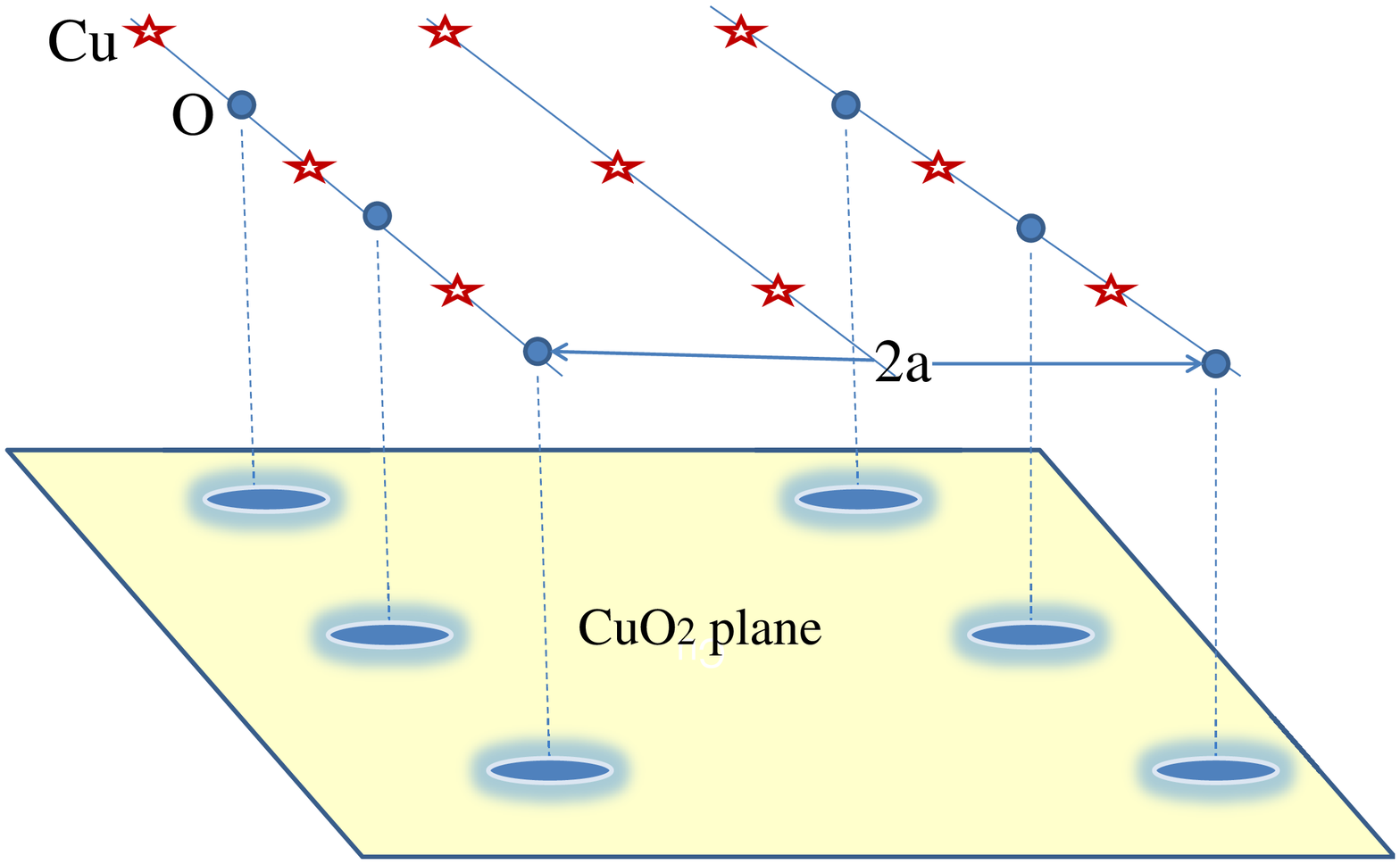}
\includegraphics[angle=-0,width=0.25\textwidth]{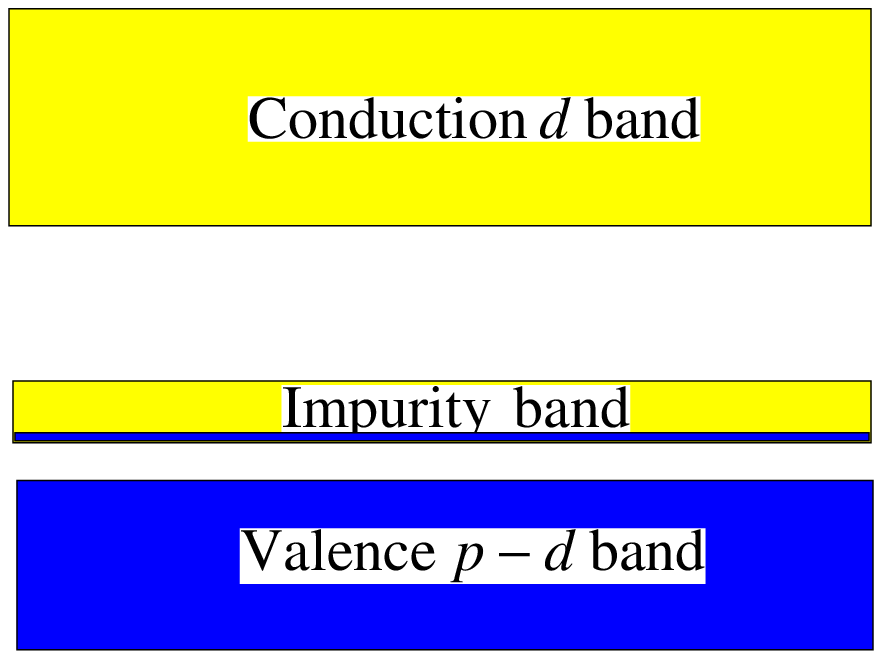}
\vskip -0.5mm \caption{(Color online) The ortho-II phase of YBa$_2$Cu$_3$O$_{6.5}$ is characterized by a periodic alternation of empty Cu and filled
Cu-O $b$-axis chains doubling the unit cell in the $a$ direction. Extra oxygen in the full chains gives rise to
an attractive potential for the holes (upper panel) creating a coherent band inside the charge-transfer gap of the Mott-Hubbard insulator (lower panel). }  \label{potential}
\end{center}
\end{figure}

In the  LDA picture the  extra oxygen in the full chains of YBa$_2$Cu$_3$O$_{6.5}$, Fig.(\ref{potential}) splits  the planar $CuO_2$ metallic band
  in two with a gap opening at the new
Brillouin zone boundary $k_x= ±\pi/2a$,  estimated from local-density approximation calculations between 120 and 160 meV  \cite{rice}. Quite a different band structure emerges  in the charge-transfer insulator.
Our key point  is that, at variance with other doping levels, $0.5$ (per unit cell) extra oxygen   creates a coherent  band  within the charge-transfer gap rather than the localized band-tail, since  YBCO6.5 is perfectly ordered, Fig.(\ref{potential}). As shown below, this coherent band accounts for the observed quantum oscillations, and its spectral function combined with the matrix elements also explains ARPES.

 To get an insight into the coherent "impurity" band dispersion of YBCO6.5 we employ
the  LDA+U band structure of the parent insulator YBCO6.0 \cite{saw}. In such a framework,  the
$CuO_2$ in-plane states are found about 0.5 eV below the Fermi level and the first electron-removal (i.e valence) state has
a $BaO-Cu$ chain character (see Fig.5 in Ref. \cite{saw}). This valence band has its maximum at $\Gamma$ point of the Brillouin zone. Spin fluctuations and significant   electron-phonon interactions \cite{alebook} could push the in-plane states  closer to the Fermi energy, thus back into the first ionization energy range, due to polaronic level shifts missing in the LDA+U band structures. The  edge of the upper in-plane band is also found  at  $\Gamma$ point in YBCO6 \cite{saw}, therefore which  particular orbitals form the valence band  is not an issue in our  further analysis. In any case we are dealing with quasi-two dimensional carriers.

A single extra oxygen ion in the chain creates the Coulomb  attractive potential for a hole, about  $-2e^2/\epsilon \rho$, at a large enough distance $\rho$  from the ion. Solving the 2D Coulomb problem in the effective mass ($m$) approximation yields an estimate of the localized state ionization energy, $E_{im}=8m e^4/\hbar^2 \epsilon^2$. $E_{im}$ is about $87$ meV  with $m=2m_e$ measured in MQO experiments \cite{undo} and  with $\epsilon =\sqrt{\epsilon_{ab} \epsilon_c}$, where $\epsilon_{ab} =500$ \cite{erb} and  $\epsilon_c=10$ \cite {beh} are the in-plane and c-axis  dielectric permittivities, respectively,   measured in the insulating YBCO single crystals. This energy is much lower than the charge-transfer gap, which is about 1 eV \cite{saw}, so that the bound state is rather shallow.  The size of the bound-state wave function, $f_{im}(\rho)\propto \exp(-\rho/a_{im})$, is  comparable to  the lattice constant, $a_{im}= \hbar^2 \epsilon /4me^2 \approx 0.4$ nm.
\begin{figure}
\begin{center}
\includegraphics[angle=-0,width=0.44\textwidth]{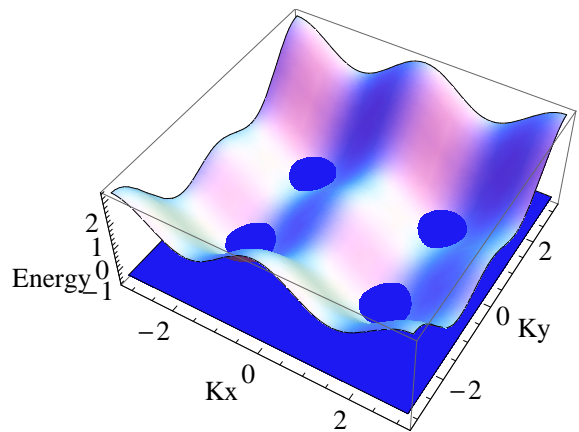}
\vskip -0.5mm \caption{(Color online) The impurity-band energy dispersion in  YBa$_2$Cu$_3$O$_{6.5}$ with four small electron pockets at the reduced Brillouin zone boundary (dark oval spots) as described by Eq.(\ref{dispersion3}) with $t^\prime=0.7 t$ and the Fermi level $\mu=-2.5t$ (Energy is measured in $2t$ and $a=1$). }  \label{dispersionfig}
\end{center}
\end{figure}

Excess oxygen ions in YBCO6.5 give rise to a potential $V({ \overrightarrow{\rho}})$,  Fig.(\ref{potential}), periodic in a plane $\overrightarrow{\rho} \equiv \{x,y\}$.  It  is  "gentle"  \cite{kohn} in the outlined sense due to the high polarizability, $\epsilon \gg 1$ of perovskites, so that  the valence-band states  of the parent insulator are mainly involved.  Hence  the impurity-band wave function, $\psi ({\bf r})$  can be expanded in a complete set of the orthogonal valence-band Wannier functions, $w({\bf r})$, which   account for most of the correlations \cite{ovc}. These functions are atomic-like with their extension, $a_0$  smaller than the lattice constant, $a$. Thus
 \begin{equation}
 \psi({\bf r})=\sum _{\bf m} F({\bf m}) w({\bf r}-{\bf m}), \label{function}
 \end{equation}
where the envelope function $F({\bf m})$ satisfies the following differential equation \cite{sub}
 \begin{equation}
 [E(-i \overrightarrow{\nabla}_{\rho})+V(\overrightarrow{\rho})] F(\overrightarrow{\rho})=E_{im} F( \overrightarrow{\rho}). \label{envelope}
 \end{equation}
Here $E(\textbf{p})$ is the LDU+U dispersion of the 2D valence band \cite{saw}.
\begin{figure}
\begin{center}
\includegraphics[angle=-0,width=0.34\textwidth]{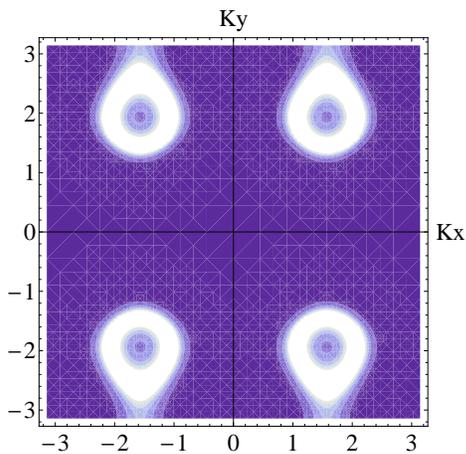}
\vskip -0.5mm \caption{(Color online) The momentum-distribution map of the impurity-band spectral function at the Fermi surface with $t^\prime=0.7 t$, $\mu=-2.5t$ and $\gamma=0.04 t$ (here $a=1$). }  \label{map}
\end{center}
\end{figure}

 The  eigenstates of Eq.(\ref{envelope}) can be expanded as
\begin{equation}
F_{\textbf{k}}(\overrightarrow{\rho})={1\over{\sqrt{N_{im}}}}\sum_{\textbf{n}}e^{i \textbf{k}\cdot \textbf{n}} f( \overrightarrow{\rho}- \textbf{n}) \label{envelope2}
\end{equation}
since the potential  $V(\overrightarrow{\rho})$ is periodic. Here $ \textbf{k}$ is the quasi-momentum in the  reduced 2D Brillouin zone, $ (-\pi/2a< k_x <\pi/2a, -\pi/a< k_y <\pi/a)$,  $ f( \overrightarrow{\rho}- \textbf{n})$ are the orthogonal impurity-band Wannier orbitals built of the bound-state wave function $f_{im}( \overrightarrow{\rho})$, and $\textbf{ n}$
are  2D position vectors of excess-oxygen ions $N_{im}$ or their projections in the plane, Fig.(\ref{potential}). Then the impurity-band dispersion is found as
\begin{equation}
 E_{im}( \textbf{k})= -\sum_{ \textbf{n}}e^{i \textbf{k}\cdot  \textbf{n}} t(\textbf{n}), \label{dispersion}
 \end{equation}
 where $ t(\textbf{n})=-\int d  \overrightarrow{\rho} f( \overrightarrow{\rho}+ \textbf{n}) [E(-i \overrightarrow{\nabla}_{\rho})+V(\overrightarrow{\rho})]f(\overrightarrow{\rho})$ are the impurity-band hopping integrals, which are positive for the attractive $V(\overrightarrow{\rho})$. It should be noted that the impurity-band dispersion, Eq.(\ref{dispersion}) holds for the whole Brillouin zone as long as we do not expand the valence-band energy operator $E(-i \overrightarrow{\nabla})$ to any finite order in $\overrightarrow{\nabla}$. The envelope function Eq.(\ref{envelope}) works correctly to any power in $k$ as long as the extension of the valence-band Wannier orbitals is much less than the extension of the impurity-band orbitals, $a_0\ll a_{im}\approx a$.
\begin{figure}
\begin{center}
\includegraphics[angle=-0,width=0.34\textwidth]{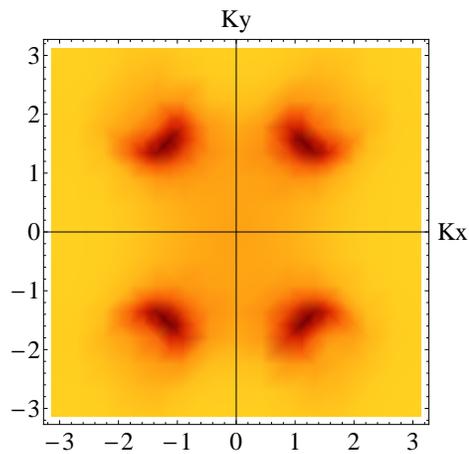}
\includegraphics[angle=-0,width=0.27\textwidth]{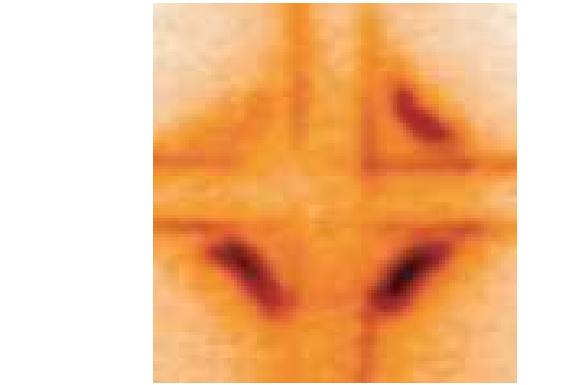}
\vskip -0.5mm \caption{(Color online) The ARPES momentum-distribution map of the impurity-band (upper panel),  Eq.(\ref{arpes}) with $(a_{im}/a)^2=0.4$, $t^\prime=0.7 t$ and $\gamma=0.94 t$. Lower panel: the experimental map of  the Fermi surface in YBCO6.5 \cite{hos} (here $a=1$).}  \label{arpesfig}
\end{center}
\end{figure}

One can parametrise the impurity-band dispersion by keeping nearest, $t$ and next nearest, $t^\prime$,  hopping integrals in Eq.(\ref{dispersion}). Thus
\begin{equation}
 E_{im}^h(\textbf{k})= - 2t\cos (k_y a)-2t^\prime [\cos (2k_x a)+\cos (2k_y a)] , \label{dispersion2}
 \end{equation}
for holes with the lower band-edge at  $\Gamma$ point, which is compatible with the LDA+U band structure of YBCO6.0 \cite{saw}. Every extra oxygen donates two holes, so that the impurity band is almost full with holes in YBCO6.5 while some holes are still localized by residual disorder in the chains and bound into preformed pairs (bipolarons) \cite{alebook,alearpes,aletun}. The electron impurity-band dispersion,
\begin{equation}
E_{im}^e(\textbf{k})=2 t\cos (k_y a)+2 t^\prime [\cos (2k_x a)+\cos (2k_y a)] , \label{dispersion3}
\end{equation}
has its minima  at the boundary of the reduced Brillouin zone, $\textbf{k}^\star =(\pm \pi/2a,  \pm [\pi-\cos^{-1}(t/2t^\prime)]/a)$ accounting for  small electron pockets in MQO \cite{undo}, Fig.(\ref{dispersionfig}). Placing the Fermi level near the bottom of the electron impurity band, $\mu=-2.5 t$, yields the size of the electron Fermi pocket about 2$\%$ of the Brillouin zone as observed \cite{undo}. There is some anisotropy in the  effective electron mass in the pocket,  $m_x=\hbar^2/8 t^\prime a^2$ and $m_y=\hbar^2/(8t^\prime-3t^2/t^\prime)a^2$. Taking $m_x=2m_e$ \cite{undo}, and $t^\prime=0.7 t$ provides a reasonable estimate for $t^\prime \approx 32$ meV and $t \approx 46$ meV.

A  momentum-distribution map of the impurity-band spectral function at the Fermi surface, $A(\textbf{k}, 0) \propto 1/[(E_{im}^e(\textbf{k})-\mu)^2+\gamma^2]$  is  shown in Fig.(\ref{map}) with the inverse quasiparticle life-time, $\gamma=0.04t$ about the Dingle temperature measured in MQO experiments \cite{undo}. It  hardly resembles the observed ARPES map \cite{hos}, which is not surprising because  the impurity-band ARPES matrix element, $M(\textbf{K})$ strongly depends on the photoelectron momentum $\textbf{K}$ \cite{alearpes}.  Calculating $M(\textbf{K})\propto \int d \textbf{r} \exp (i \textbf{K} \cdot \textbf{r}) \psi(\textbf{r})$
with the impurity-band wave function Eq.(\ref{function}) one obtains $M(\textbf{K}) \propto f_{\textbf{K}_\parallel}$, where $f_{\textbf{k}}$ is the Fourier component of the impurity-band Wannier orbital, $f( \overrightarrow{\rho})$, introduced in Eq.(\ref{envelope2}). The extension of this orbital is comparable to  the lattice constant, which explains the strong dependence of the matrix element on the in-plane  photoelectron momentum ${\textbf{K}_\parallel}$.

Approximating $f( \overrightarrow{\rho})$ as the 2D Coulomb bound state $f_{im}(\rho)$ yields
\begin{equation}
M(\textbf{K}) \propto {1\over{[1+(K_{\parallel} a_{im})^2]^{3/2}}},
\end{equation}
and the ARPES momentum-distribution map of the Fermi surface, $I(\textbf{K},0) \propto M(\textbf{K})^2A(\textbf{K}, 0)$, is
\begin{equation}
I(\textbf{K},0) \propto {1\over{[1+(K_{\parallel} a_{im})^2]^{3} [(E_{im}^e({\textbf{K}_\parallel})-\mu)^2+\gamma^2]}}. \label{arpes}
\end{equation}

 A very close resemblance between  theoretical Eq.(\ref{arpes})  and  experimental ARPES  maps  is shown in Fig.(\ref{arpesfig}). The theory reproduces the locations, relative intensities and the half-moon shape of the ARPES spots in YBCO6.5 reconciling  puzzling ARPES and MQO data. In fact, the small electron pockets observed in MQO experiments \cite{undo} are seen in ARPES \cite{hos},  partially shadowed by the matrix element.
 The ARPES quasiparticle life-time  turns out  much shorter than in MQO, presumably  due to an instrumental broadening and a high  level of impurity scattering off deposited potassium atoms on cleaved YBCO \cite{hos}. In the electron-doped cuprates, one expects an impurity band somewhat below the conduction band, if these compounds are doped charge-transfer insulators as the hole-doped ones.  Then  the outlined scenario  with reversed holes and electrons also accounts for the small \textit{hole} pockets  observed in MQO of the \textit{electron}-doped cuprates \cite{kar}.

 I gratefully acknowledge A. Bansil, A. M. Bratkovsky, V. V. Kabanov, M. V. Kartsovnik, Y. V. Kopelevich, R. S. Markiewicz,  D. Mihailovic, S. G. Ovchinnikov, and L. Taillefer  for illuminating discussions. This work was supported  by the European Union Framework Programme 7 (NMP3-SL-2011-263104-HINTS) and by the Visiting Professor Program of Unicamp (Campinas, Brasil).

\end{document}